\documentclass[a4paper,11pt]{article}

\usepackage{amsmath}
\usepackage{amssymb}
\usepackage{color}
\usepackage{cite}
\usepackage{hyperref}

\makeatletter
\@addtoreset{equation}{section}
\renewcommand{\theequation}{\thesection.\@arabic\c@equation}
\makeatother
\makeatletter
\renewcommand\appendix{\par
  \setcounter{section}{0}%
  \setcounter{subsection}{0}%
  \gdef\thesection{Appendix \@Alph\c@section }
  \renewcommand{\theequation}
  {\Alph{section}.\arabic{equation}}
}
\makeatother

\def \be {\begin{equation}}
\def \ee {\end{equation}}
\def \ba {\begin{array}}
\def \ea {\end{array}}
\def \bea{\begin{eqnarray}}
\def \eea{\end{eqnarray}}

\def \a {\alpha}

\def \G {\Gamma}

\def \m {\mu}
\def \n {\nu}

\def \l {\lambda}
\def \L {\Lambda}
\def \s {\sigma}

\def \r {\rho}
\def \o {\omega}
\def \O {\Omega}
\def \th {\theta}
\def \Th {\Theta}
\def \t {\tau}

\def \p {\partial}
\def \f {\frac}
\def \na {\nabla}

\def \nn {\nonumber}
\def \scl {\ell}
\def \ma {\mathcal}

\def \lt {\left}
\def \rt {\right}

\def \sr {\sqrt}
\def \td {\tilde}
\def \hs {\hspace}
\def \pp {\propto}

\def \Im {{\textrm{Im}}}

\title{\textbf{RN/CFT Correspondence From Thermodynamics}}
\author{
Bin Chen$^{1,2}$\footnote{bchen01@pku.edu.cn}\,
and
Jia-ju Zhang$^{1}$\footnote{jjzhang@pku.edu.cn}
}
\date{}

\begin{document}

\maketitle

\begin{center}
{\it
$^{1}$Department of Physics and State Key Laboratory of Nuclear Physics and Technology, Peking University, Beijing 100871, P.R. China\\
\vspace{2mm}
$^{2}$Center for High Energy Physics, Peking University, Beijing 100871, P.R. China\\
}
\vspace{10mm}
\end{center}

\begin{abstract}

Recent studies suggest that in the Kerr/CFT correspondence, much universal information of the dual CFT, including the central charges and the temperatures, is fully encoded in the thermodynamics of the outer and inner horizons of the Kerr(-Newman) black holes. In this paper, we study holographic descriptions of Reissner-Nordstr\"om (RN) black holes in arbitrary dimensions by using the thermodynamics method.We refine the thermodynamics method proposed in  \cite{Chen:2012mh} by imposing the ``quantization'' condition so that we can fix the ambiguity in determining the central charges of the dual CFT of RN black holes. Using the refined thermodynamics method, we find the holographic CFT duals for the RN black holes, and confirm these pictures by using conventional analysis of asymptotic symmetry group and the hidden conformal symmetry in the low-frequency scattering. In particular, we revisit the four-dimensional dyonic RN black hole and find a novel magnetic picture, besides the known electric CFT dual picture. We show how to generate a class of dual dyonic pictures by $SL(2,Z)$ transformations.

\end{abstract}

\baselineskip 18pt
\thispagestyle{empty}

\newpage

\section{Introduction}

The Kerr/CFT correspondence  \cite{K1} asserts that there is a two-dimensional (2D) conformal field theory (CFT)  to describe the Kerr black hole holographically. In setting up the Kerr/CFT correspondence, the conventional way is to obtain the central charges of dual CFT from the asymptotic symmetry group (ASG) of near-horizon geometry of extreme black hole in  either Barnich-Brandt-Compere (BBC) formalism  \cite{K1,BBC} or equivalently the stretched horizon formalism  \cite{Carlip1,Carlip2,Chen:2011wm}, and read the dual temperatures from the Frolov-Thorne vacuum  \cite{K1} or the hidden conformal symmetry in the low-frequency scattering  \cite{K2}. Kerr/CFT has many extensions and generalizations, and the reader can find details and more complete references in the nice reviews \cite{R1,R2}.

One remarkable feature in the holographic description of Kerr and multi-charged black holes is that the central charges of the dual CFT are written in terms of ``quantized" charges, angular momenta and $U(1)$ charges, independent of the mass of the black holes. This feature could be related to the fact that the area product of the horizons $S_+ S_-$ of these black holes are also mass-independent. Actually, it was shown \cite{Cvetic:1996kv,Larsen:1997ge,Cvetic:1997uw,Cvetic:1997xv} that for general five-dimensional (5D) and four-dimensional (4D) multi-charged rotating black holes, the outer and inner horizon entropies could be written respectively as
\be
S_\pm=2\pi (\sr{N_L} \pm \sr{N_R}), \label{entropypm}
\ee
where $N_L,N_R$ could be interpreted as the levels of the left- and right-moving sectors in a two-dimensional CFT. Therefore the entropy product
\be
S_+S_-=4\pi^2(N_L-N_R)
\ee
should be quantized, as $ (N_L-N_R)$ must be integer due to the level matching condition in CFT. As a result, the entropy product $S_+S_-$ must be mass-independent \cite{Cvetic:2009jn,Cvetic:2010mn}. For other recent relevant studies on this issue, see \cite{Castro:2012av,Visser:2012zi,Detournay:2012ug, Visser:2012wu}. Strictly speaking, the mass-independence condition breaks down in some cases, including various warped black holes in three-dimensional (3D) topologically massive gravity, but the relation (\ref{entropypm}) is always sound for the black holes with holographic descriptions  \cite{Detournay:2012ug}. From (\ref{entropypm}), one may find microscopical entropy of dual CFT. This suggests that the physics of the inner horizon of the black hole should be taken seriously.

Very recently, the Kerr/CFT correspondence was investigated from the point of view of thermodynamics of both outer and inner horizons \cite{Chen:2012mh}.   Firstly, it was proved that the first law of thermodynamics of the outer horizon always indicates that of the inner horizon, under reasonable assumption. Secondly, the mass-independence of the entropy product $S_+S_-$ is equivalent to the condition $T_+S_+=T_-S_-$, which is much easier to check. More interestingly, it was found that the thermodynamics in the left- and right-moving sectors of the dual CFT could be obtained from the linear composition of the thermodynamics of the outer and inner horizons  \cite{Cvetic:2009jn, Chen:2012mh}. This thermodynamics method  provides us a simple way to read the information of the dual CFT. It has been checked in many well-established black hole/CFT correspondences, including 3D BTZ, 4D Kerr-Newman and 5D Myers-Perry black holes \cite{Strominger:1997eq,Lu:2008jk,Hartman:2008pb,Wang:2010qv,Chen:2010xu,Krishnan:2010pv,Chen:2010ywa,Chen:2011wm,Chen:2011kt}, and applied to the study of holographic descriptions of black rings \cite{Chen:2012yd}. It turns out to be quite effective, allowing us to read the central charges and the temperatures in all possible pictures.

One of interesting generalizations of Kerr/CFT is the so-called RN/CFT correspondence \cite{Hartman:2008pb,Garousi:2009zx,Chen:2009ht}, which states that there is a holographic 2D CFT description for the four-dimensional Reissner-Nordstr\"om (RN) black hole.  The central charges of dual CFT have been computed either from a reduced two-dimensional effective gravity action or from a uplifted 5D metric point of view \cite{Chen:2009ht}. It is puzzling to see that the central charge could only be determined up to a scale factor $c=6Q^3/l$, with $l$ being an undetermined factor. Correspondingly, there seems to be an one-parameter class of CFTs dual to 4D RN black hole. Such an ambiguity looks strange if one apply the same techniques to the well-known multi-charged black holes in string theory, whose CFT duals have quantized central charges proportional to the product of the numbers of different branes. We try to solve this puzzle in this paper.
The key point in our treatment is to impose the ``quantization" condition on the thermodynamics method, which allows us to get rid of the ambiguity. We find that this condition is actually in accord with the quantization condition on the angular momentum of the higher dimensional uplifted configuration.

Another interesting issue in the RN/CFT correspondence is the holographic duals for dyonic RN black holes. It has been studied using the hidden conformal symmetry in  \cite{Chen:2010yu}.  In  \cite{Chen:2010yu}, the dual picture was obtained by using an electric-charged scalar to probe the geometry. This picture will be called  as the electric ($E$) picture. As the dyonic RN black hole carries both electric and magnetic charges, it is interesting to inquire what one can get if using a magnetic charged probe. We will show that such an investigation gives us a magnetic dual picture of the dyonic RN black hole. Actually, the magnetic ($M$) picture could be easier to figure out from the thermodynamics method, as we will show in section~\ref{s4}. As shown in the case of Kerr-Newman black hole, when the black hole has  two $U(1)$ symmetries, there will be a CFT dual picture for every $U(1)$ charge  \cite{Hartman:2008pb,Chen:2010ywa,Wang:2010qv,Chen:2010xu,Chen:2010yu}, and a whole class of novel CFT pictures could be generated by  $SL(2,Z)$ transformations acting on two elementary dual pictures \cite{Chen:2011wm,Chen:2011kt}. We find that the similar phenomenon happens for 4D dyonic RN black holes. In this case, there is an electromagnetic duality group $SL(2,Z)$ acting on the elementary electric and magnetic pictures.

In this paper we investigate RN/CFT correspondence mainly using the thermodynamics method, and verify our results using conventional methods if possible.  In Section~\ref{s2} we consider RN black holes in all dimensions $d\geq4$, and find that $T_+S_+=T_-S_-$ are always satisfied.  We find their CFT duals using the thermodynamics method, and verify the pictures by re-deriving the results via ASG analysis and the hidden conformal symmetry. In Section~\ref{s3}, we consider RN-AdS black holes in all dimensions, and find that $T_+S_+=T_-S_-$ breaks down, which suggests that there are no CFT duals for such black holes. In Section~\ref{s4}, we consider the four-dimensional dyonic RN black hole, and find a novel magnetic CFT dual. The picture is confirmed by the study of hidden conformal symmetry in low frequency scattering of various kinds of probe scalar and also ASG analysis of a 6D uplifted spacetime. In Section~\ref{s6}, we end with conclusion and discussion.

\section{RN/CFT in arbitrary dimensions}\label{s2}

In this section we consider the RN/CFT correspondence in spacetime of dimension $d \geq 4$. We set up the general RN/CFT in three different ways, i.e.\! the thermodynamics method, ASG analysis, and the hidden conformal symmetry.

\subsection{Black hole solutions}

The charged spherically symmetric black hole solutions in $d$ dimensions were found in  \cite{Myers:1986un}. We have $c=\hbar=1$ for convenience, but we set $G_d=\ell_p^{d-2}$ with $\ell_p$ being the Planck length in $d$-dimensional spacetime. We use the convention here because the dimensional analysis plays a subtle role in our calculation. We consider the Einstein-Maxwell theory with the action
\be
I_d=\f{1}{16\pi G_d}\int d^d x \sr{-g}R-\f{1}{4\O_{d-2}}\int d^d x \sr{-g}F_{\m\n}F^{\m\n},
\ee
where we have normalized the electromagnetic field so that $\O_{d-2}$ is the volume of a unit $d-2$ sphere $S^{d-2}$
\be
\O_{d-2}=\f{2\pi^{\f{d-1}{2}}}{\G(\f{d-1}{2})}.
\ee
Note that in four dimensions we are using the Gauss convention with this action. The Einstein equation is now
\bea
&& R_{\m\n}-\f{1}{2}R g_{\m\n}=8\pi G_d T_{\m\n},  \nn\\
&& T_{\m\n}=\f{1}{\O_{d-2}} \lt( F_{\m\r}F_{\n}^{\phantom{\n}\r}-\f{1}{4}g_{\m\n}F_{\r\s}F^{\r\s} \rt).
\eea

The $d$-dimensional RN black hole has the metric and the electromagnetic potential
\bea \label{e2}
&& ds^2_d=-N^2 dt^2+g_{rr}dr^2+r^2 d\O_{d-2}^2, \nn\\
&& A=-\f{Q}{(d-3)r^{d-3}}dt,
\eea
with
\bea
&& N^2=\f{1}{g_{rr}}=1-\f{2m}{r^{d-3}}+\f{q^2}{r^{2(d-3)}},  \nn\\
&& Q^2=\f{(d-2)(d-3)\O_{d-2}}{8\pi G_d}q^2.
\eea
The black hole has outer and inner horizons locating at $r_\pm$, with
\be
r_\pm^{d-3}=m\pm\sr{m^2-q^2}.
\ee
The mass of the black hole, the Hawking temperatures and the entropies of the outer and inner horizons are respectively
\bea
&& M=\f{(d-2)\O_{d-2}}{8\pi G_d}m=\f{(d-2)\O_{d-2}}{16\pi \ell_p^{d-2}}(r_+^{d-3}+r_-^{d-3}),  \nn\\
&& T_\pm=\lt|\f{\p_r N^2}{4\pi}\rt|_{r=r_\pm}=\f{(d-3)(r_+^{d-3}-r_-^{d-3})}{4\pi r_\pm^{d-2}}, \\
&& S_\pm=\f{A_\pm}{4 G_d}=\f{\O_{d-2}r_\pm^{d-2}}{4 \ell_p^{d-2}}. \nn
\eea
The electric charge of the black hole is $Q$, which is just
\be
Q^2=\f{(d-2)(d-3)\O_{d-2}}{8\pi \ell_p^{d-2}}(r_+r_-)^{d-3},
\ee
and the electric potentials at the outer and inner horizons are
\be
\Phi_\pm=\f{Q}{(d-3)r_\pm^{d-3}}.
\ee
One can see that in $d$ dimensions the electric charge has the dimension of length power $\f{d-4}{2}$, i.e.\! $[Q]=\ma L^{\f{d-4}{2}}$. It is dimensionless in four dimensions but is not so in higher dimensions.

One can verify the first laws of thermodynamics of the outer and inner horizons
\bea
&& d M=T_+  d S_++\Phi_+ d Q  \nn\\
&& \phantom{dM}=-T_-d S_-+\Phi_- d Q,
\eea
which are equivalent to Smarr formulas of the two horizons
\bea
&& M=\f{d-2}{d-3}T_+S_+ +\Phi_+ Q  \nn\\
&& \phantom{M}=-\f{d-2}{d-3}T_-S_- +\Phi_- Q.
\eea

\subsection{Refined thermodynamics method} \label{s2.2}

Let us apply the thermodynamics method proposed in  \cite{Chen:2012mh}. First of all, we check that $T_+S_+=T_-S_-$, which means that the entropy product $S_+S_-$ is independent of the mass $M$ and there is CFT dual for the black hole with equal right- and left-moving central charges. We define the new quantities \cite{Cvetic:1997uw,Cvetic:1997xv,Cvetic:2009jn}
\bea \label{e10}
&& T_{R,L}=\f{T_+T_-}{T_- \pm T_+},  \nn\\
&& S_{R,L}=\f{1}{2}(S_+ \mp S_-),  \\
&& \Phi_{R,L}=\f{T_-\Phi_+ \pm T_+\Phi_-}{2(T_- \pm T_+)},\nn
\eea
such that the first laws and the Smarr formulas can be separated into the right- and left-moving sectors
\bea \label{e1}
&& \f{1}{2} d M=T_R d S_R+\Phi_R  d Q  \nn\\
&& \phantom{\f{1}{2}dM}=T_L d S_L+\Phi_L d Q,
\eea
\bea
&& \f{1}{2}M=\f{d-2}{d-3}T_R S_R+\Phi_R Q  \nn\\
&& \phantom{\f{1}{2}M}=\f{d-2}{d-3}T_L S_L+\Phi_L Q.
\eea
Explicitly, these quantities are
\bea
&& T_{R,L}=\f{(d-3)(r_+^{d-3}- r_-^{d-3})}{4\pi(r_+^{d-2}\pm r_-^{d-2})},  \nn\\
&& S_{R,L}=\f{\O_{d-2}}{8 \ell_p^{d-2}}(r_+^{d-2}\mp r_-^{d-2}),  \\
&& \Phi_{R,L}=\f{r_+ \pm r_-}{4(r_+^{d-2}\pm r_-^{d-2})}\sr{\f{(d-2)\O_{d-2}(r_+r_-)^{d-3}}{2\pi(d-3) \ell_p^{d-2}}}.\nn
\eea
We have to mention that, the Smarr formulas play no fundamental rule in our calculations and they are just convenient ways to verify the first laws.

As what have been done in  \cite{Chen:2012mh,Chen:2012yd} for the BTZ black hole, 4D Kerr-Newman black hole, 5D Meyers-Perry black hole, doubly rotating and dipole black rings, one could rewrite the first laws of black hole as
\be \label{e36}
d J=T_L^J d S_L-T_R^J  d S_R,
\ee
then one identify $T_{R,L}^J$ as the $J$ picture CFT temperatures. Since the angular momentum $J$ and the entropies $S_{R,L}$ are dimensionless, the CFT temperatures $T_{R,L}^J$ are dimensionless as required. This treatment works remarkably well to get the $J$ pictures of various black holes with rotations.

For the RN black holes, we may use the same strategy. From (\ref{e1}) we can get
\be \label{e37}
dQ=\f{T_L}{\Phi_R-\Phi_L}dS_L-\f{T_R}{\Phi_R-\Phi_L}dS_R.
\ee
But now the left hand side of the equation is not dimensionless, and so a factor must be multiplied to both sides of the equation. In  \cite{Chen:2012mh}, we multiplied an arbitrary factor to read the $Q$ picture of the Kerr-Newman black hole and found complete agreement with the results in the literature  \cite{Chen:2009ht}.
However, such an ambiguity makes us uncomfortable, especially considering the fact that for multi-charged black holes in string theory this treatment may give us bizarre results. Actually without dimension uplifting or reducing, the only natural scale is the Planck length ${1}/\ell_p^{\f{d-4}{2}}$, but still there is an ambiguity in introducing a numerical dimensionless factor.

Note that in setting up the $J$ picture, we should work with Eq. (\ref{e36}). The underlying reason is simple, as the angular momentum is quantized. In fact,  Eq. (\ref{e36}) tells us how the black hole responds to the perturbation. As the angular momentum is quantized, the minimal variation due to the perturbation gives exactly Eq. (\ref{e36}).

Similarly, for the equation (\ref{e37}), the left hand side should be quantized. It is more suggestive to recast the first laws into the form
\be \label{e41}
d N=T_L^N d S_L -T_R^N d S_R,
\ee
with $N$ being an integer-valued quantized charge, then the temperatures of the $N$ picture CFT dual are $T_{R,L}^N$ without any ambiguity. For the RN black holes,  we scale (\ref{e37}) as
\be \label{e40}
\f{\l}{\ell_p^{\f{d-4}{2}}}dQ=\f{\l T_L}{\ell_p^{\f{d-4}{2}}(\Phi_R-\Phi_L)}dS_L-\f{\l T_R}{\ell_p^{\f{d-4}{2}}(\Phi_R-\Phi_L)}dS_R,
\ee
where $\l$ is a numerical factor. The factor $\l$ makes
\be \label{e38} \f{\ell_p^{\f{d-4}{2}}}{\l}=e \ee
with $e$ being the unit charge which is determined by the Maxwell theory. Then we can identify the size of the space where the two-dimensional CFT reside as  \cite{Cvetic:1997uw,Cvetic:1997xv,Cvetic:2009jn}
\be
R_Q=\f{\l}{\ell_p^{\f{d-4}{2}}(\Phi_R-\Phi_L)}
   =\f{4\pi(d-3)(r_+^{2(d-2)}-r_-^{2(d-2)})\l\ell_{p}}{(r_+^{d-3}-r_-^{d-3})\sr{2\pi(d-2)(d-3)\O_{d-2}(r_+ r_-)^{d-1}}},
\ee
and the temperatures of the CFT as
\be \label{e5}
T_{R,L}^Q=R_Q T_{R,L}=\f{(d-3)^2(r_+^{d-2} \mp r_-^{d-2})\l\ell_{p}}{\sr{2\pi(d-2)(d-3)\O_{d-2}(r_+ r_-)^{d-1}}}.
\ee
We suppose that the right- and left-moving entropies could be expressed in the form of the Cardy formula if there really exists a CFT dual
\be
S_{R,L}=\f{\pi^2}{3}c_{R,L}^Q T_{R,L}^Q,
\ee
then we obtain the right- and left-moving central charges
\bea \label{e4}
&& c_{R,L}^Q=\f{3S_{R,L}}{\pi^2T_{R,L}^Q}=\f{3}{4\l\ell_p^{d-1}}\sr{\f{(d-2)\O_{d-2}^3(r_+r_-)^{d-1}}{{2\pi^3(d-3)^3}}} \nn\\
&& \phantom{c_{R,L}^Q}
            =\f{3}{(d-3)^2\l}\lt( \f{4}{(d-2)(d-3)} \rt)^{\f{1}{d-3}} \lt( \f{\O_{d-2}}{2\pi} \rt)^{\f{d-4}{d-3}}
             \lt( \f{Q}{\ell_p^{\f{d-4}{2}}} \rt)^{\f{d-1}{d-3}}.
\eea

For example, in four-dimensions we at last get
\be
c_{R,L}^Q=\f{6Q^3}{\l},
\ee
with the numerical factor $\l=1/e$. Therefore, we resolve the puzzle on the undetermined scale factor in RN/CFT  \cite{Chen:2009ht}.

If the Maxwell theory is that in quantum electrodynamics (QED), then the unit charge $e$ is related to the fine structure constant as
\be
\a=e^2\simeq \f{1}{137}.
\ee
Since the black hole charge is also quantized, we have the black hole $Q=Ne$ with $N$ being a possibly very large integer. Then the central charges become
\be
c_{R,L}^Q={6 \a^2 N^3}.
\ee
The appearance of the fine structure constant in the central charge is actually in accord with the discussion in \cite{Visser:2012zi}.

Furthermore the first laws (\ref{e1}) could be written in a more suggestive way
\bea
&& T_R^Q d S_R=R_Q \lt( \f{1}{2}dM-\Phi_R dQ \rt),  \nn\\
&& T_L^Q d S_L=R_Q \lt( \f{1}{2}dM-\Phi_L dQ \rt).
\eea
Under some perturbations $dM=\o$, $dQ=k_e e$, with $e$ being the unit charge (\ref{e38}) and so $k_e$ being an integer, we identify
\bea
&& T_R^Q d S_R=\o_R^Q-q_R^Q\m_R^Q,  \nn\\
&& T_L^Q d S_L=\o_L^Q-q_L^Q\m_L^Q,
\eea
with $\o_{R,L}^Q$, $q_{R,L}^Q$, $\m_{R,L}^Q$ being the frequencies, the charges, and the  chemical potentials of the perturbation around the thermodynamical equilibrium of finite-temperature CFT. Explicit calculations show that
\bea \label{e9}
&& \o_{R}^Q=\o_{L}^Q=\f{R_Q}{2}\o
         =\f{2\pi(d-3)(r_+^{2(d-2)}-r_-^{2(d-2)})\l\ell_{p}\o}{(r_+^{d-3}-r_-^{d-3})\sr{2\pi(d-2)(d-3)\O_{d-2}(r_+ r_-)^{d-1}}}, \nn\\
&& q_{R}^Q=q_{L}^Q=k_e,  \nn\\
&& \m_{R}^Q=e R_Q\Phi_{R}=\f{(r_++r_-)(r_+^{d-2}-r_-^{d-2})}{2r_+r_-(r_+^{d-3}-r_-^{d-3})},  \\
&& \m_{L}^Q=e R_Q\Phi_{L}=\f{(r_+-r_-)(r_+^{d-2}+r_-^{d-2})}{2r_+r_-(r_+^{d-3}-r_-^{d-3})}. \nn
\eea
These results will be compared with those obtained from the hidden conformal symmetry.

\subsection{ASG analysis}

To do ASG analysis we uplift the $d$-dimensional RN black hole to $(d+1)$-dimensional Einstein gravity
\be
I_{d+1}=\f{1}{16\pi G_{d+1}}\int d^{d+1}x \sr{-G}R_{d+1}.
\ee
The metric becomes
\be \label{e3}
ds^2_{d+1}=ds^2_d+\f{16\pi}{\O_{d-2}} \lt( \ell_{d+1}d\chi+\ell_p^{\f{d-2}{2}}A \rt)^2,
\ee
with $ds^2_d$ and $A$ being defined as (\ref{e2}), and $\chi\sim\chi+2\pi$ and $\ell_{d+1}$ being the scale of the extra dimension. Again the natural scale of $\ell_{d+1}$ is the Planck length $\ell_p$ up to some numerical constant $\ell_{d+1}=\l \ell_p$. From Kaluza-Klein reduction, we have
\be
R_{d+1}=R_d-\f{4\pi\ell_p^{d-2}}{\O_{d-2}}F_{\m\n}F^{\m\n},
\ee
and thus we have the identification of two theories $I_{d+1}=I_d$ with
\be
G_{d+1}=\sr{\f{16\pi}{\O_{d-2}}}2\pi\l\ell_{p}G_d.
\ee
For uplifted RN black hole, the areas of the horizons satisfy
\be
(A_\pm)_{d+1}=\sr{\f{16\pi}{\O_{d-2}}}2\pi\l\ell_{p}(A_\pm)_{d},
\ee
 so that we always have the relationships
\be
\f{(A_\pm)_{d+1}}{G_{d+1}}=\f{(A_\pm)_{d}}{G_{d}}=\f{A_{\pm}}{\ell_p^{d-2}},
\ee
where we have $G_{d}=\ell_p^{d-2}$ and denote $A_\pm=(A_\pm)_{d}$. The uplifting does not change the entropies of the black holes. Note also that,  the electric charge $Q$ has been transformed to the angular momentum along the angle $\chi$. The angular momentum and angular velocities of the outer and inner horizons could be identified as
\be \label{e62}
J_\chi=\f{\l}{\ell_p^{\f{d-4}{2}}}Q, ~~~ \O_\pm^\chi=\f{\ell_p^{\f{d-4}{2}}}{\l}{\Phi_\pm}.
\ee
It is remarkable that the quantization of the angular momentum  $J_\chi$ indicates the relationship
\be \label{e39} \f{\ell_p^{\f{d-4}{2}}}{\l}=e \ee
with $e$ being the unit charge of the theory. In other words, the quantization condition imposed on the thermodynamics method in the last subsection is equivalent to the quantization of the angular momentum in the uplifted spacetime. The requirement of a quantized angular momentum which is crucial to pin down the extra factor in the central charge has been ignored in the literature.

To do ASG analysis, we take the extremal limit of the metric ($\ref{e3}$). We expand the following quantities at the horizon $r_+=r_-$,
\bea
&& N^2=(r-r_+)^2 f_1^2+\ma O(r-r_+)^3,  \nn\\
&& g_{rr}=\f{f_2^2}{(r-r_+)^2}+\ma O\lt(\f{1}{r-r_+}\rt),  \\
&& N^\chi=\f{\ell_p^{\f{d-4}{2}}}{\l} {A_t}=-\O_+^\chi+(r-r_+)f^\chi_3+\ma O(r-r_+)^2.  \nn
\eea
Explicitly, we have
\be
f_1=\f{d-3}{r_+}, ~~~ f_2=\f{r_+}{d-3}, ~~~ f_3^\chi=\f{\ell_p^{\f{d-4}{2}}Q}{\l r_+^{d-2}},
\ee
and we may also define
\be
f^\chi\equiv \f{f_2 f_3^\chi}{f_1}=\f{\ell_p^{\f{d-4}{2}}Q}{(d-3)^2\l r_+^{d-4}}.
\ee
Remember that in the extremal case, we have the areas of the horizons (in $d$-dimensional spacetime) and the electric charge
\be
A_\pm=\O_{d-2}r_\pm^{d-2}, ~~~ Q= \sr{\f{(d-2)(d-3)\O_{d-2}}{8\pi\ell_p^{d-2}}} r_+^{d-3}.
\ee
It was demonstrated in  \cite{Carlip2,Chen:2011wm} that for an extremal black hole, there is always a CFT dual, whose information could be read from ASG analysis, no matter in the BBC formalism  \cite{BBC}, or in the stretched horizon formalism \cite{Carlip1,Carlip2}. The left-moving central charge of the CFT is
\bea
&& c_L^\chi=\f{3f^\chi A_+}{2\pi G_d}=\f{3r_+^{d-1}}{4\l\ell_{p}^{d-1}}\sr{\f{(d-2)\O_{d-2}^3}{{2\pi^3(d-3)^3}}} \nn\\
&& \phantom{c_L^\chi}
            =\f{3}{(d-3)^2\l}\lt( \f{4}{(d-2)(d-3)} \rt)^{\f{1}{d-3}} \lt( \f{\O_{d-2}}{2\pi} \rt)^{\f{d-4}{d-3}}
             \lt( \f{Q}{\ell_p^{\f{d-4}{2}}} \rt)^{\f{d-1}{d-3}}.
\eea
And from the Frolov-Thorne vacuum the left-moving temperature of the CFT is read out
\be
T_L^\chi=\f{1}{2\pi f^\chi}=\f{2(d-3)^2\l \ell_{p}}{r_+\sr{2\pi(d-2)(d-3)\O_{d-2}}}.
\ee
Comparing the results with (\ref{e4}), (\ref{e5}), we see that in the extremal limit $c_L^Q=c_L^\chi$ and $T_L^Q=T_L^\chi$, the results are in perfect match.  Especially, the results justify the factor multiplied in (\ref{e40}) and thus the prescription (\ref{e41}). This shows that the thermodynamics method is an effective way of getting the CFT dual of the black hole.

\subsection{Hidden conformal symmetry}

We investigate the scattering of a complex scalar off the RN black hole.  We can consider either a charged scalar in the $d$-dimensional RN black hole background (\ref{e2}), or equivalently a neutral scalar in the uplifted $(d+1)$-dimensional black hole  background (\ref{e3}). In the former case, we consider a scalar of mass $\m_d$ and charge $k_e e$, with $e$ being the unit charge (\ref{e38}) and $k_e$ being an integer. The equation of motion for such scalar $\Phi$ is
\be \label{e6}
(\na_\m-i k_e e A_\m)(\na^\m-i k_e e A^\m)\Phi=\m_d^2\Phi.
\ee
We define $\r=r^{d-3}$, and expand $\Phi=e^{-i\o t}R(\r)\Th_\L$, with $\Th_\L$ being the eigenfunction of the Laplace operator of the unit $d-2$ sphere $S^{d-2}$, i.e.\! ($D^i D_i+\L)\Th_\L=0$. Then we could get the radial euqation
\be \label{e8}
\p_\r(\r-\r_+)(\r-\r_-)\p_\r R(\r)+\f{r^2 \lt( \o\r-\f{Q k_e e}{d-3} \rt)^2}{(d-3)^2(\r-\r_+)(\r-\r_-)}R(\r)=\f{\L+m_d^2 r^2}{(d-3)^2}R(\r).
\ee

In the later case we consider a scalar of mass $\m_{d+1}$, with its equation of motion being
\be \label{e7}
\na_M \na^M \Phi=\m_{d+1}^2\Phi.
\ee
We expand $\Phi=e^{-i\o t+ik_\chi\chi}R(\r)\Th_\L$, and then it can be shown that the equations (\ref{e6}) and (\ref{e7}) are identical as long as we have (\ref{e39}) and
\be \label{e63}
k_e=k_\chi, ~~~ \m_d^2=\m_{d+1}^2+\f{\O_{d-2}k_\chi^2}{16\pi \l^2 \ell_{p}^2}.
\ee
It is crucial that the integer $k_e$ is identified with the integer $k_\chi$, whose quantization is due to the periodic nature of $\chi$. The importance of this consistent identification has been ignored in the literature.

Under some suitable approximations in the low-frequency limit, from the radial equation (\ref{e8}), one can arrive at
\be \label{e38}
\p_\r(\r-\r_+)(\r-\r_-)\p_\r R(\r)+\f{r_+^2 \lt( \o\r_+-\f{Q k_e e}{d-3} \rt)^2}{(d-3)^2(\r_+-\r_-)(\r-\r_+)}R(\r)
-\f{r_-^2 \lt( \o\r_--\f{Q k_e e}{d-3} \rt)^2}{(d-3)^2(\r_+-\r_-)(\r-\r_-)}R(\r)
=K R(\r),
\ee
with $K$ being some constant.

In the study of hidden conformal symmetry for the non-extreme black hole, the conformal coordinates could be defined as\footnote{ The hidden conformal symmetry in four and five dimensional rotating and multi-charged black holes in string theory has been firstly discussed in \cite{Cvetic:1997uw,Cvetic:1997xv}.} \cite{K2}
\bea
&&\o^+=\sr{\f{\r-\r_+}{\r-\r_-}}e^{2 \pi T_R^C \psi +2n_R^C t},  \nn\\
&&\o^-=\sr{\f{\r-\r_+}{\r-\r_-}}e^{2 \pi T_L^C \psi +2n_L^C t},  \nn\\
&&y=\sr{\f{\r_+-\r_-}{\r-\r_-}}e^{\pi (T_R^C+T_L^C) \psi +(n_R^C+n_L^C )t}.
\eea
Here $t$ is the time, $\r$ is the radial coordinate which is not necessarily $r$ but can be a monotonically increasing function of $r$, and $\psi \sim \psi+2\pi$  may be an angle of the spacetime, or an internal angle, or a supposition of some angles. Also we use the letter $C$ to denote $CFT$, $T_{R,L}^C$ are the right- and left-moving central charges of the CFT, and $n_{R,L}^C$ have no immediate physical meaning now. With the conformal coordinates the vector fields could be locally defined as
\bea
&&H_1=\p_+, \nn\\
&&H_0= \o^+\p_++\frac{1}{2}y\p_y , \nn\\
&&H_{-1}=\o^{+2}\p_++\o^+y\p_y-y^2\p_-,
\eea
and
\bea
&&\tilde H_1=\p_- \nn\\
&&\tilde H_0=\o^-\p_-+\frac{1}{2}y\p_y \nn\\
&&\tilde H_{-1}=\o^{-2}\p_-+\o^-y\p_y-y^2\p_+.
\eea
These vector fields obey the $SL(2, R)$ Lie algebra
\be
[H_0, H_{\pm 1}]=\mp H_{\pm 1},\hs{5ex} [H_{1},H_{-1}]=2H_0,
\ee
and similarly for $(\tilde H_0, \tilde H_{\pm 1})$.

The quadratic Casimir is
\bea
&&\ma H^2=\tilde{\ma H}^2=H_0^2-\frac{1}{2}(H_1 H_{-1}+H_{-1}H_1) \nn\\
&&\phantom{\ma H^2}=\frac{1}{4}(y^2\p^2_y-y\p_y)+y^2 \p_+\p_-.
\eea
In terms of $(t,\r,\psi)$ coordinates, the Casimir becomes
\bea
&&\ma H^2=\p_\r(\r-\r_+)(\r-\r_-)\p_\r
-\f{(\r_+-\r_-)[\pi(T_L^C+T_R^C)\p_t-(n_L^C+n_R^C)\p_\psi]^2}{16\pi^2 (T_L^C n_R^C-T_R^C n_L^C)^2 (\r-\r_+)} \nn\\
&&\phantom{\ma H^2=} +\f{(\r_+-\r_-)[\pi(T_L^C-T_R^C)\p_t-(n_L^C-n_R^C)\p_\psi]^2}{16\pi^2 (T_L^C n_R^C-T_R^C n_L^C)^2 (\r-\r_-)}.
\eea
With the scalar field being expanded as $\Phi=e^{-i\o t+i k \psi}R(\r)$, the equation $\ma H^2\Phi=K\Phi$ gives us the radial equation of motion
\bea \label{e39}
&&\p_\r(\r-\r_+)(\r-\r_-)\p_\r R(\r)
+\f{(\r_+-\r_-)[\pi(T_L^C+T_R^C)\o+(n_L^C+n_R^C)k]^2}{16\pi^2 (T_L^C n_R^C-T_R^C n_L^C)^2 (\r-\r_+)} R(\r) \nn\\
&&-\f{(\r_+-\r_-)[\pi(T_L^C-T_R^C)\o+(n_L^C-n_R^C)k]^2}{16\pi^2 (T_L^C n_R^C-T_R^C n_L^C)^2 (\r-\r_-)} R(\r)=K R(\r),
\eea
where $K$ is a constant. Note that $k$ is the quantum number along the angle $\psi$ and must be integer-valued.

Identifying the above two radial equations (\ref{e38}), (\ref{e39}), we find
\bea
&& k=k_e,  \nn\\
&& T_{R,L}^Q=\f{(d-3)^2(r_+^{d-2} \mp r_-^{d-2})\l\ell_{p}}{\sr{2\pi(d-2)(d-3)\O_{d-2}(r_+ r_-)^{d-1}}},  \nn\\
&& n_{R,L}^Q=-\f{(d-3)(r_+ \mp r_-)}{4r_+ r_-}.
\eea
The temperatures obtained here are in perfect accord with the results got from the thermodynamics method (\ref{e5}). Here, the remarkable point is that the integer-valued quantum number $k$ is consistently identified with the integer-valued charge $k_e$. This point has not been taken seriously in the former study of RN/CFT. It helps us to pin down the ambiguous factor in the central charge of RN/CFT. Again, the result justifies the prescription (\ref{e41}).

The above  radial equation can be solved in terms of hypergeometric functions and gives the retarded Green's function and the absorption cross section that agree with the predictions of the CFT side \cite{arXiv:1001.3208, Chen:2010xu},
\bea
&&G_R \pp \sin \lt(\pi h_L^Q+ i \f{ \o_L^Q-q_L^Q \m_L^Q}{2T_L^Q} \rt)
                          \sin \lt(\pi h_R^Q+ i \f{ \o_R^Q-q_R^Q\m_R^Q}{2T_R^Q} \rt)  \nn\\
&&\phantom{G_R\pp} \times \G \lt( h_L^Q-i \f{\o_L^Q-q_L^Q\m_L^Q}{2\pi T_L^Q} \rt)
                            \G \lt( h_L^Q+i \f{\o_L^Q-q_L^Q\m_L^Q}{2\pi T_L^Q} \rt)  \nn\\
&&\phantom{G_R\pp} \times\G \lt( h_R^Q-i \f{\o_R^Q-q_R^Q\m_R^Q}{2\pi T_R^Q} \rt)
                            \G \lt( h_R^Q+i \f{\o_R^Q-q_R^Q\m_R^Q}{2\pi T_R^Q} \rt),
\eea
\bea
&&\s \pp \sinh \lt(\f{ \o_L^Q-q_L^Q \m_L^Q}{2T_L^Q}+\f{ \o_R^Q-q_R^Q\m_R^Q}{2T_R^Q} \rt)  \nn\\
&&\phantom{\s\pp} \times  \lt| \G \lt( h_L^Q+i \f{\o_L^Q-q_L^Q\m_L^Q}{2\pi T_L^Q} \rt) \rt|^2
                           \lt| \G \lt( h_R^Q+i \f{\o_R^Q-q_R^Q\m_R^Q}{2\pi T_R^Q} \rt) \rt|^2.
\eea
The conformal weights, the frequencies, the charges, and the chemical potentials of the perturbations in the CFT side could be identified as
\bea
&& h_{R,L}^Q=\f{1}{2}\pm\sr{\f{1}{4}+K},  \nn\\
&& \o_{R,L}^Q=\f{\pi T_L^Q T_R^Q \o}{T_L^Q n_R^Q-T_R^Q n_L^Q}
             =\f{2\pi(d-3)(r_+^{2(d-2)}-r_-^{2(d-2)})\l\ell_{p}\o}{(r_+^{d-3}-r_-^{d-3})\sr{2\pi(d-2)(d-3)\O_{d-2}(r_+ r_-)^{d-1}}}, \nn\\
&& q_{R,L}^Q=k_e, \\
&& \m_R^Q=-\f{T_R^Q n_L^Q}{T_L^Q n_R^Q-T_R^Q n_L^Q}=\f{(r_++r_-)(r_+^{d-2}-r_-^{d-2})}{2r_+r_-(r_+^{d-3}-r_-^{d-3})}, \nn\\
&& \m_L^Q=-\f{T_L^Q n_R^Q}{T_L^Q n_R^Q-T_R^Q n_L^Q}=\f{(r_+-r_-)(r_+^{d-2}+r_-^{d-2})}{2r_+r_-(r_+^{d-3}-r_-^{d-3})}.   \nn
\eea
These quantities  got from the hidden conformal symmetry are in perfect match with the ones got in the thermodynamics methods (\ref{e9}). This agreement is remarkable. On one hand, the thermodynamics of the black hole tells us how it respond to the perturbation. On the other hand, the scatting amplitude of the probe scalar gives us the information of the black hole. It is amazing to see that the thermodynamics method gives us almost the same information on the dual CFT as the probe scalar: the same frequencies, the charges and the chemical potentials.

\section{RN-AdS black holes in arbitrary dimensions}\label{s3}

The general RN-AdS black holes in arbitrary dimensions were found in  \cite{Xu:1988ju}, and the solutions are similar to their asymptotically flat cousins. The theory has the action
\be
I_d=\f{1}{16\pi G_d}\int d^d x \sr{-g}(R-2\L)-\f{1}{4\O_{d-2}}\int d^d x \sr{-g}F_{\m\n}F^{\m\n},
\ee
with the Newton constant $G_d=\ell_d^{d-2}$ the cosmological constant
\be
\L=-\f{(d-1)(d-2)}{2\scl^2}.
\ee
The equation of motion becomes
\bea
&& R_{\m\n}-\f{1}{2}R g_{\m\n}+\L g_{\m\n}=8\pi G_d T_{\m\n},  \nn\\
&& T_{\m\n}=\f{1}{\O_{d-2}} \lt( F_{\m\r}F_{\n}^{\phantom{\n}\r}-\f{1}{4}g_{\m\n}F_{\r\s}F^{\r\s} \rt).
\eea

The $d$-dimensional RN-AdS black hole has the metric and the electromagnetic potential
\bea
&& ds^2_d=-N^2 dt^2+g_{rr}dr^2+r^2 d\O_{d-2}^2, \nn\\
&& A=-\f{Q}{(d-3)r^{d-3}}dt,
\eea
with
\bea
&& N^2=\f{1}{g_{rr}}=1-\f{2m}{r^{d-3}}+\f{q^2}{r^{2(d-3)}}+\f{r^2}{\ell^2},  \nn\\
&& Q^2=\f{(d-2)(d-3)\O_{d-2}}{8\pi G_d}q^2.
\eea
From $N^2(r_\pm)=0$, we can get the location of the outer and inner horizons $r_\pm$, and represent the parameters of the black hole $m, q$ in terms of $r_\pm$. The mass of the black hole, the Hawking temperatures and the entropies of the outer and inner horizons are respectively
\bea
&& M=\f{(d-2)\O_{d-2}}{8\pi G_d}m,  \nn\\
&& T_\pm=\lt|\f{\p_r N^2}{4\pi}\rt|_{r=r_\pm}, \\
&& S_\pm=\f{A_\pm}{4 G_d}=\f{\O_{d-2}r_\pm^{d-2}}{4\ell^{d-2}}. \nn
\eea
The electric charge of the black hole is $Q$, and the electric potentials at the outer and inner horizons are
\be
\Phi_\pm=\f{Q}{(d-3)r_\pm^{d-3}}.
\ee
One can verify the first laws of the outer and inner horizons
\bea
&& d M=T_+  d S_++\Phi_+ d Q  \nn\\
&& \phantom{dM}=-T_-d S_-+\Phi_- d Q.
\eea
Note that there are no trivial Smarr formulas here.

For example, in four dimensions, we have
\bea
&& M=\f{r_++r_-}{2\ell_p^2}(1+\f{r_+^2+r_-^2}{2\scl^2}),  \nn\\
&& T_+=\f{r_+-r_-}{4\pi r_+^2}(1+\f{3r_+^2+2r_+r_-+r_-^2}{\ell^2}), \nn\\
&& T_-=\f{r_+-r_-}{4\pi r_-^2}(1+\f{r_+^2+2r_+r_-+3r_-^2}{\ell^2}), \nn\\
&& S_\pm=\f{\pi r_\pm^2}{\ell_p^2},  \\
&& \Phi_+=\f{1}{\ell_p}\sr{\f{r_-}{r_+}(1+\f{r_+^2+r_+r_-+r_-^2}{\ell^2})}, \nn\\
&& \Phi_-=\f{1}{\ell_p}\sr{\f{r_+}{r_-}(1+\f{r_+^2+r_+r_-+r_-^2}{\ell^2})}, \nn\\
&& Q=\f{1}{\ell_p}\sr{{r_+}{r_-}(1+\f{r_+^2+r_+r_-+r_-^2}{\ell^2})}. \nn
\eea
The first laws can be verified easily, and we can see the symmetry of the quantities under the exchange of $r_\pm$ proposed in \cite{Chen:2012mh}. But now we have
\be
T_+S_+-T_-S_-=\f{(r_++r_-)(r_+-r_-)^2}{2\ell_p^2\ell^2},
\ee
which is not vanishing. This suggests that the entropy product $S_+S_-$ is mass-dependent and there should be no CFT dual for the four-dimensional RN-AdS black hole\footnote{The mass-dependence of $S_+S_-$ of four-dimensional RN-AdS black hole has been observed in \cite{Visser:2012wu}.}. For the RN-AdS black holes in higher dimensions, we have similar conclusion. In five dimensions, we get
\be
T_+S_+-T_-S_-=\f{\pi(r_+^2-r_-^2)^2}{4\ell_p^3\ell^2}.
\ee
We check that for $d=6\sim30$, we always have $T_+S_+-T_-S_-$ nonvanishing, and we believe that the result holds in all higher dimensions. As a result, we conclude that there seems to be no CFT dual for the RN-AdS black holes in all dimensions. Actually, one may naively using the thermodynamics method proposed in Section 2 to the RN-AdS case, as now the first laws at both horizons are still well-defined. But one would find that the central charges of left-moving and right-moving sectors are different. The result contradicts with our expectation since   in Einstein gravity without diffeomorphism anomaly the central charges in both sectors of candidate CFT should be the same, leading to $T_+S_+=T_-S_-$ \cite{Chen:2012mh}.

\section{Four-dimensional Dyonic RN black holes}\label{s4}

Four-dimensional RN black hole is special compared to its higher dimensional cousins in the sense that it not only can carry electric charge but can also carry magnetic charge, namely  in four dimensions there are electromagnetically charged, i.e.\! dyonic, RN black hole. We investigate the holographic descriptions of the dyonic black hole using the thermodynamics method and the hidden conformal symmetry. We find that there are two elementary CFT duals, namely the known electric ($E$) picture  \cite{Chen:2010yu} and a novel magnetic ($M$) picture, from which the other dyonic pictures could be generated by $SL(2,Z)$ transformations. Since the embedding of the dyonic RN black hole in higher dimensions is nontrivial, we cannot use the ASG formalism in a straightforward way. However, as we show, the dyonic black hole geometry could be understood as the solution of a theory with two $U(1)$ fields, which allows us to uplift the solutions to six dimensions and analyze its ASG.

\subsection{Dyonic RN black hole}

The dyonic RN black hole is a solution of the action
\be
S=\f{1}{16\pi G_4} \int d^4x\sr{-g}R-\f{1}{16\pi} \int d^4x\sr{-g}F_{\m\n}F^{\m\n}
  -\f{\th e^2}{32\pi^2} \int d^4x\sr{-g}F_{\m\n}*F^{\m\n},
\ee
with $*$ being Hodge duality. Here to discuss the full $SL(2,Z)$ symmetry, we have introduced the $\th$-term for the gauge field. The two real constants $e,\th$ could be combined into a complex coupling parameter
\be
\t=\f{\th}{2\pi}+\f{i}{e^2}.
\ee
Again, we use the convention $c=\hbar=1$ and $G_4=\ell_p^2$, and for the electromagnetic part we have used the Gauss convention. The metric of the dyonic RN black hole is of the form
\be \label{e18}
ds^2_4=-\lt( 1-\f{2 G_4 M}{r}+\f{G_4 Q^2}{r^2} \rt)dt^2+\lt( 1-\f{2 G_4 M}{r}+\f{ G_4 Q^2}{r^2} \rt)^{-1}dr^2+r^2 (d\th^2+\sin^2\th d\phi^2).
\ee
Here $M$ is the mass of the black hole, and
\be
Q^2=Q_e^2+Q_m^2,
 \ee
with $Q_{e,m}$ being the electric and magnetic charges of the black hole respectively. The gauge field of the theory can be written as
\be \label{e12}
A=-\f{Q_e}{r}dt+Q_m (\cos\th \mp 1) d\phi.
\ee
The upper minus sign applies to the sphere with the south pole deleted, say $0 \leq \th <\pi$, and the lower plus sign applies to the sphere with the north pole deleted, say $0 < \th \leq \pi$. We denote $F=dA$ and $*F=d\td A$,
then we have the dual electromagnetic potential
\be \label{e13}
\td A=-\f{Q_m}{r}dt-Q_e (\cos\th \mp 1) d\phi.
\ee

Due to the Witten effect  \cite{Witten:1979ey}, the electric and magnetic charges are respectively
\bea \label{qeqm}
&& Q_e=N_e e-N_m \f{e\th}{2\pi},  \nn\\
&& Q_m=\f{N_m}{e},
\eea
with $N_{e,m}$ being integers. Note that for two dyons with charges $Q_{e,m}$ and $Q_{e',m'}$, there should be the Dirac-Zwanziger-Schwinger quantization condition
\be \label{dzs}
Q_e Q_{m'}-Q_m Q_{e'}=N_e N_{m'}-N_m N_{e'} \in Z.
\ee

The horizons locate at $r_\pm=G_4M\pm\sr{G_4^2M^2-G_4Q^2}$, and the temperatures, the entropies, the electric and magnetic potentials of the outer and inner horizons are respectively
\bea
&& T_\pm=\f{r_+ - r_-}{4\pi r_\pm^2}, \nn\\
&& S_\pm=\f{\pi r_\pm^2}{\ell_p^2},  \nn\\
&& \Phi_\pm^{e,m}=\f{Q_{e,m}}{r_\pm}.
\eea
There are the first laws at the outer and inner horizons
\bea \label{e45}
&& dM=T_+ dS_++\Phi_+^e dQ_e +\Phi_+^m dQ_m \nn\\
&& \phantom{dM}=-T_- dS_-+\Phi_-^e dQ_e +\Phi_-^m dQ_m ,
\eea
which are equivalent to the Smarr formulas
\bea
&& M=2T_+ S_++\Phi_+^e Q_e +\Phi_+^m Q_m \nn\\
&& \phantom{M}=-2T_- S_-+\Phi_-^e Q_e +\Phi_-^m Q_m .
\eea

\subsection{Thermodynamics method}\label{s4.2}

According to the discussion in Sect 2, $Q_{e,m}$ are not good quantum numbers and we should use the integers $N_{e,m}$ that appear in (\ref{qeqm}) to apply the thermodynamics method. We rewrite the first laws (\ref{e45}) as
\bea \label{e50}
&& dM=T_+ dS_++\O_+^e dN_e +\O_+^m d N_m \nn\\
&& \phantom{dM}=-T_- dS_-+\O_-^e d N_e +\O_-^m d N_m ,
\eea
with $N_{e,m}$ being the integers appear in (\ref{qeqm}) and
\be
\O_\pm^e=\f{e Q_e}{r_\pm}, ~~~ \O_\pm^m=\f{\f{Q_m}{e}-e Q_e\f{\th}{2\pi}}{r_\pm}.
\ee
With the quantities defined as those in (\ref{e10}),
\bea
&& T_{R,L}=\f{T_+T_-}{T_- \pm T_+},  \nn\\
&& S_{R,L}=\f{1}{2}(S_+ \mp S_-),  \nn\\
&& \O_{R}^{e,m}=\f{T_-\O_+^{e,m} + T_+\O_-^{e,m}}{2(T_- + T_+)},\nn\\
&& \O_{L}^{e,m}=\f{T_-\O_+^{e,m} - T_+\O_-^{e,m}}{2(T_- - T_+)},
\eea
 the first laws could be recast into
\bea \label{e11}
&& \f{1}{2} d M=T_R d S_R+\O_R^e  d N_e+\O_R^m  d N_m  \nn\\
&& \phantom{\f{1}{2}dM}=T_L d S_L+\O_L^e  d N_e+\O_L^m  d N_m,
\eea
with
\bea
&& T_R=\f{r_+-r_-}{4\pi(r_+^2+r_-^2)}, ~~~ T_L=\f{1}{4\pi(r_++r_-)},  \nn\\
&& S_{R,L}=\f{\pi}{2\ell_p^2}(r_+^2 \mp r_-^2), \nn\\
&& \O_R^{e}=\f{e Q_{e}(r_++r_-)}{2(r_+^2+r_-^2)}, ~~~ \O_L^{e}=\f{e Q_{e}}{2(r_+-r_-)}, \nn\\
&& \O_R^{m}=\f{\lt(\f{Q_m}{e}-e Q_e\f{\th}{2\pi}\rt)(r_++r_-)}{2(r_+^2+r_-^2)}, ~~~
   \O_L^{m}=\f{\f{Q_m}{e}-e Q_e\f{\th}{2\pi}}{2(r_+-r_-)}.
\eea

Setting $dN_m=0$ in (\ref{e11}), we could get
\be
dN_e=\f{T_L}{\O_R^e-\O_L^e}dS_L-\f{T_R}{\O_R^e-\O_L^e}dS_R,
\ee
then we may identify the scale factor and the temperatures of underlying CFT in the $N_e$ picture, or electronic ($E$) picture as
\bea
&& R_e=\f{1}{\O_R^e-\O_L^e}=\f{(r_++r_-)(r_+^2+r_-^2)}{\ell_p^2 e Q_e Q^2},   \nn\\
&& T_{R,L}^e=R_e T_{R,L}=\f{r_+^2 \mp r_-^2}{4\pi \ell_p^2 e Q_e Q^2}.
\eea
From the Cardy formula we read the central charges
\be \label{ce}
c_{R,L}^e=\f{3S_{R,L}}{\pi^2T_{R,L}^e}={6eQ_eQ^2}.
\ee
We see that the central charges and the temperatures of CFT in the $E$ picture agree with the ones found in  \cite{Chen:2010yu} up to the overall factor we fix here. Moreover from the first laws (\ref{e11}), we may set $dM=\o, dN_e=k_e, dN_m=0$ and get
\bea
&& T_R^e d S_R=\o_R^e-q_R^e\m_R^e,  \nn\\
&& T_L^e d S_L=\o_L^e-q_L^e\m_L^e,
\eea
with the frequencies, the charges, and the chemical potentials of the perturbations in the electric picture CFT being identified as
\bea \label{e17}
&& \o_{R,L}^e=\f{R_e}{2}\o=\f{(r_++r_-)(r_+^2+r_-^2)}{2 \ell_p^2e Q_eQ^2}\o,  \nn\\
&& q_{R,L}^e=k_e,  \nn\\
&& \m_{R}^e={R_e\O_{R}^e}=\f{(r_++r_-)^2}{2 \ell_p^2 Q^2},\\
&& \m_{L}^e={R_e\O_{L}^e}=\f{r_+^2+r_-^2}{2 \ell_p^2 Q^2}. \nn
\eea

Similarly we can set $dN_e=0$ in (\ref{e11}) and get the CFT dual in the $N_m$ picture, or magnetic ($M$) picture. The scalar factor and the temperatures of the $M$ picture CFT could be identified as
\bea \label{tm}
&& R_m=\f{1}{\O_R^m-\O_L^m}=\f{(r_++r_-)(r_+^2+r_-^2)}{ \ell_p^2 Q^2 \lt(\f{Q_m}{e}-e Q_e\f{\th}{2\pi} \rt) },   \nn\\
&& T_{R,L}^m=R_m T_{R,L}=\f{r_+^2 \mp r_-^2}{4\pi\ell_p^2 Q^2 \lt( \f{Q_m}{e}-e Q_e\f{\th}{2\pi} \rt)},
\eea
and the central charges are
\be \label{cm}
c_{R,L}^m=\f{3S_{R,L}}{\pi^2T_{R,L}^m}={6Q^2 \lt( \f{Q_m}{e}-e Q_e\f{\th}{2\pi} \rt)}.
\ee
Furthermore, the frequencies, the charges, and the chemical potentials of the perturbation in the magnetic picture CFT, which is dual to the perturbation $dM=\o,dN_e=0,dN_m=k_m$ in the gravity side, could be identified as
\bea \label{oqmm}
&& \o_{R,L}^m=\f{R_m}{2}\o=\f{(r_++r_-)(r_+^2+r_-^2)}{2 \ell_p^2Q^2 \lt( \f{Q_m}{e}-e Q_e\f{\th}{2\pi} \rt) }\o,  \nn\\
&& q_{R,L}^m=k_m,  \nn\\
&& \m_{R}^m={R_m\O_{R}^m}=\f{(r_++r_-)^2}{2 \ell_p^2 Q^2},\\
&& \m_{L}^m={R_m\O_{L}^m}=\f{r_+^2+r_-^2}{2 \ell_p^2 Q^2}. \nn
\eea

Note that we call the electronic and magnetic pictures as $N_{e,m}$ pictures or $E,M$ pictures, and try to avoid the name $Q_{e,m}$ pictures, because  $Q_{e.m}$ are not good quantum numbers and they are not integers, but $N_{e,m}$ are integers. Now we have two CFT pictures for 4D dyonic RN black hole, namely the $E$ and $M$ pictures. From our experience in 4D Kerr-Newman and 5D Myers-Perry black holes, once there are two dual pictures, there could be a class of dual pictures related by $SL(2,Z)$ transformations with each other  \cite{Chen:2011wm,Chen:2011kt}. In these cases, the $SL(2,Z)$ group could be understood as the T-duality group. In the case of dyonic RN black hole, we may try to generate more dual pictures from $SL(2,Z)$ transformations as well. Using the description in  \cite{Chen:2012mh}, we redefine the charges $N_{e,m}$ and their intensive quantities $\O_\pm^{e,m}$ as
\bea \label{e51}
&&
\lt( \ba{c} N_{e'} \\ N_{m'} \ea \rt)=
\lt(\ba{cc} d & -c \\ -b & a \ea \rt)
\lt( \ba{c} N_e \\ N_m \ea \rt),  \nn\\
&&
\lt( \ba{c} \O_\pm^{e'} \\ \O_\pm^{m'} \ea \rt)=
\lt(\ba{cc} a & b \\ c & d \ea \rt)
\lt( \ba{c} \O_\pm^{e} \\ \O_\pm^{m}  \ea \rt),
\eea
with
\be
\lt(\ba{cc} a & b \\ c & d \ea \rt) \in SL(2,Z),
\ee
and so
\be
\lt(\ba{cc} d & -c \\ -b & a \ea \rt) \in SL(2,Z).
\ee
We stress that the justification of the redefinitions is that the charges $N_{e,m}$ are integers. Then the first laws (\ref{e50}) become
\bea
&& dM=T_+ dS_++\O_+^{e'} dN_{e'} +\O_+^{m'} d N_{m'} \nn\\
&& \phantom{dM}=-T_- dS_-+\O_-^{e'} d N_{e'} +\O_-^{m'} d N_{m'}.
\eea
From the first laws, a pair of generic dyonic pictures with temperatures $T_{R,L}^{e',m'}$ and central charges $c_{R,L}^{e',m'}$ could be obtained as
\bea \label{e28}
&&\lt( \ba{c} 1/T_{R,L}^{e'}\\1/T_{R,L}^{m'} \ea \rt)=
\lt(\ba{cc} a & b \\ c & d \ea \rt)
\lt( \ba{c} 1/T_{R,L}^{e}\\1/T_{R,L}^{m} \ea \rt),  \nn\\
&&
\lt( \ba{c} c_{R,L}^{e'} \\ c_{R,L}^{m'} \ea \rt)=
\lt(\ba{cc} a & b \\ c & d \ea \rt)
\lt( \ba{c} c_{R,L}^{e}\\c_{R,L}^{m} \ea \rt).
\eea
We show in \cite{Chen:2012pt} that this $SL(2,Z)$ symmetry originates from the electromagnetic duality in the four-dimensional Einstein-Maxwell theory.

There is a simple way to understand various pictures in the thermodynamics method. The thermodynamics of the black hole tells us how the black hole responds with respect to the perturbations of the infalling particle carrying the mass and the charges. The electric and magnetic charges of the perturbation $q_{e,m}$ could be expressed in terms of two integers $k_{e,m}$ as suggested in (\ref{qeqm})
\bea \label{qeqm1}
&& q_e=k_e e-k_m \f{e\th}{2\pi},  \nn\\
&& q_m=\f{k_m}{e}.
\eea
If the perturbation carries only an electric charge, or more accurately $k_m=0$, then the thermodynamics laws tell us how the right- and left-moving sectors changes with the charges. This gives us the electric picture of the black hole (\ref{ce}). While if the perturbation carries only a magnetic charge, or more accurately $k_e=0$, the thermodynamics laws tell us the magnetic picture (\ref{cm}). If the perturbation carries both the electric and magnetic charges, the thermodynamics laws give us the dyonic pictures (\ref{e28}).

On the other hand, if we consider a probe scattering off the RN black hole, its scattering amplitude encodes the information of the black hole as well. As we will show in next subsection, if the probe is electrically (magnetically) charged, then it tells us the electric (magnetic) picture. While if the probe is dyonic, then it gives us a dyonic picture.

\subsection{Hidden conformal symmetry}

The hidden conformal symmetry of an electrically charged scalar scattering off the dyonic RN black hole was considered in \cite{Chen:2010yu}, from which the electric CFT dual was found. Here we consider the scattering of a more general dyonic charged scalar, and try to find other  CFT duals. Suppose that there is a complex scalar with the electric and magnetic charges (\ref{qeqm1}) and mass $\m$. Its equation of motion is just
\be
(\na_\m-i q_e A_\m-i q_m \td A_\m)(\na^\m-i q_e A^\m-i q_m \td A^\m)\Phi=\m^2\Phi,
\ee
with $A,\td A$ defined as in (\ref{e12}) and (\ref{e13}). Note that the coupling of the magnetic charge with the background could be determined from the electromagnetic duality. Now the scalar has to be expanded as $\Phi=e^{-i\o t+i[k_\phi \mp (Q_m q_e-Q_e q_m)]\phi}R(r)\Th(\th)$ with upper sign applying to the north pole and the lower sign applying to the south pole \cite{Semiz:1991kh}. Note that the scalar picks a factor $e^{i2(Q_m q_e-Q_e q_m)\phi}$ when passing from north to south, and Dirac-Zwanziger-Schwinger quantization condition (\ref{dzs}) could make the factor single-valued.

The equation of motion could be decomposed into the angular part and the radial part
\be
\f{1}{\sin\th}\p_\th\sin\th\p_\th \Th(\th)-\f{[k_\phi-(Q_m q_e-Q_e q_m)\cos\th]^2}{\sin^2\th}\Th(\th)+\L\Th(\th)=0
\ee
\be\label{e26}
\p_r(r-r_+)(r-r_-)\p_r R(r)+\f{r^2[r\o-(Q_e q_e+Q_m q_m)]^2}{(r-r_+)(r-r_-)}R(r)=(\L+\m^2r^2) R(r),
\ee
with $\L$ as the separation constant. Since the black hole is static, the quantum number $k_\phi$ does not appear in the radial equation. Under the conditions of low frequency, small mass, small electric and magnetic charges, and near region approximations \cite{Chen:2010yu}, the radial equation could be written as
\bea \label{e14}
&& \p_r(r-r_+)(r-r_-)\p_r R(r)+\f{r_+^2[r_+\o-(Q_e q_e+Q_m q_m)]^2}{(r_+-r_-)(r-r_+)}R(r)  \nn\\
&&  -\f{r_-^2[r_-\o-(Q_e q_e+Q_m q_m)]^2}{(r_+-r_-)(r-r_-)}R(r)=K R(r),
\eea
with $K$ being some constant. Note that from (\ref{qeqm}) and (\ref{qeqm1}), in the above radial equation there is
\be
Q_e q_e+Q_m q_m=e Q_e k_e + \lt( \f{Q_m}{e}-e Q_e\f{\th}{2\pi} \rt)k_m.
\ee

To get the electric picture CFT, we set the magnetic charge of the probe scalar vanishing in (\ref{e14}), which means $k_m=0$ not simply $q_m=0$.  Comparing the radial equations (\ref{e14}) with  (\ref{e39}), we find
\be
k=k_e, ~~~ T_{R,L}^e=\f{r_+^2 \mp r_-^2}{4\pi \ell_p^2 Q^2 e Q_e},  ~~~ n_{R,L}^e=-\f{r_+ \mp r_-}{4 \ell_p^2 Q^2}.
\ee
The temperatures are exactly the same as the ones found in the thermodynamics method. Just like in Section~\ref{s2}, we could get the retarded Green's function and the absorption cross section  that agree with the those of the CFT. The conformal weights, the frequencies, the charges, and the chemical potentials of the perturbation in the CFT could be identified as
\bea
&& h_{R,L}^e=\f{1}{2}\pm\sr{\f{1}{4}+K},  \nn\\
&& \o_{R,L}^e=\f{\pi T_L^e T_R^e \o}{T_L^e n_R^e-T_R^e n_L^e}
             =\f{(r_++r_-)(r_+^2+r_-^2)}{2 \ell_p^2 Q^2 e Q_e}\o, \nn\\
&& q_{R,L}^e=k_e, \\
&& \m_R^e=-\f{T_R^e n_L^e}{T_L^e n_R^e-T_R^e n_L^e}=\f{(r_++r_-)^2}{2\ell_p^2Q^2},  \nn\\
&& \m_L^e=-\f{T_L^e n_R^e}{T_L^e n_R^e-T_R^e n_L^e}=\f{r_+^2+r_-^2}{2\ell_p^2Q^2}.   \nn
\eea
The results here got from the hidden conformal symmetry are in perfect agreement with the ones got in the thermodynamics methods (\ref{e17}). Therefore, we see that the scattering of the probe scalar with the electric charge gives exactly the same electric picture as in the thermodynamics method.

On the other hand, we may consider the probe scalar carrying only a magnetic charge. In this case, setting the electric charge of the scalar vanishing $k_e=0$, not $q_e=0$, in (\ref{e14}), the radial equation could still be compared with the Casimir (\ref{e39}). In this way, we find the magnetic picture with
\be
k=k_m, ~~~
T_{R,L}^m=\f{r_+^2 \mp r_-^2}{4\pi \ell_p^2 Q^2 \lt( \f{Q_m}{e}-e Q_e\f{\th}{2\pi} \rt)},  ~~~
n_{R,L}^m=-\f{r_+ \mp r_-}{4 \ell_p^2 Q^2}.
\ee
The temperatures are just (\ref{tm}). The conformal weights, the frequencies, the charges, and the chemical potentials of the perturbation in the CFT could be got as well. They are in perfect agreement with the ones got in the thermodynamics methods in the previous subsection (\ref{oqmm}).

Furthermore, we can consider the probe dyonic scalar with both the electric and magnetic charges, and find the dyonic picture suggested before. The procedure is that we make the redefinition in the radial equation (\ref{e14})
\be
\lt( \ba{c} k_{e'}\\ k_{m'} \ea \rt)=
\lt(\ba{cc} d & -c \\ -b & a \ea \rt)
\lt( \ba{c} k_{e}\\ k_{m} \ea \rt),
\ee
Then from the redefined radial equation we could set $k_{m'}=0$ or $k_{e'}=0$ and get the generic CFT dual pictures (\ref{e28}).

In summary, we see that the CFT duals got from the hidden conformal symmetry are in perfect match with the ones got in the thermodynamics method. This further verifies the robustness of the thermodynamics method of setting up the CFT duals of black holes.

\subsection{ASG analysis}

The usual way to obtain the central charges of the CFT dual for the RN black hole is to uplift the theory to a higher dimensional gravity theory \cite{Hartman:2008pb} or reduce to a 2D effective action \cite{Chen:2009ht}. However,
 for the dyonic RN black holes, there is short of direct derivation. In  \cite{Chen:2009ht}, it was argued that the central charge in 2D effective theory should be proportional to $Q^2$, up to an undetermined factor. This is in accord with what we found in the thermodynamics method. It is true that we can uplift the dyonic RN black hole into five dimension and may obtain the central charge of the electric picture, but we cannot read the central charge of the magnetic picture in a clear way. In this subsection, we provide another way to understand this problem.

The essential point is that the spacetime (\ref{e18}) of dyonic RN black hole is also the solution of a gravity theory with two $U(1)$ gauge fields,
 \be
S=\f{1}{16\pi G_4} \int d^4x\sr{-g}R-\f{1}{16\pi } \int d^4x\sr{-g} \f{1}{\Im \t}|F_{\m\n}-\t H_{\m\n}|^2,
\ee
with
\bea \label{e25}
&& G_4=\ell_p^2, ~~~
\t=\f{\th}{2\pi}+\f{i}{e^2},\nn\\
&& F=dA, ~~~ A=-\f{N_e}{r}dt,  \nn\\
&& H=dB, ~~~
B=-\f{N_m}{r}dt.
\eea
We write the action in the analog of that of type IIB supergravity. As can be checked easily, the action is invariant classically under an $SL(2,R)$ transformation
\bea \label{e60}
&& \t'=\f{a \t +b}{ c \t+d},  \nn\\
&&
\lt( \ba{c} A' \\ B' \ea \rt)=
\lt(\ba{cc} a & b \\ c & d \ea \rt)
\lt( \ba{c} A \\ B \ea \rt),
\eea
with $a,b,c,d \in R$, $ad-bc=1$. The pair of two forms $(F,H)$ transforms under SL(2,R) the same way as the pair of one forms $(A,B)$. Upon quantization $N_{e,m}$ are integers and the $SL(2,R)$ becomes $SL(2,Z)$. Now $Q_{e,m}$  in (\ref{e18}) are just the parameters in the black hole metric and are related to the integer-valued  charges $N_{e,m}$ through
\bea \label{quan}
&& Q_e=N_e e-N_m \f{e\th}{2\pi},  \nn\\
&& Q_m=\f{N_m}{e}.
\eea
Thus the $U(1)^2$ black hole could fully pertain the properties of the dyonic black hole. The thermodynamics quantities of the outer and inner horizons, and thus the first laws, are formally identical to the ones of the  dyonic black hole. So the CFT duals, including the $E$ and $M$ pictures as well as the pictures generated by $SL(2,Z)$, of the $U(1)^2$ black hole from thermodynamics are identical with the CFT duals of the dyonic black hole. This suggests that we may understand the CFT dual duals of dyonic RN black hole from the investigation of the identified CFT duals for the $U(1)^2$ black hole.

The advantage of working with the $U(1)^2$ black hole is that it could be uplifted to six dimension in a simple way and from the uplifted metric we can do ASG analysis to read the information of dual CFTs. The uplifted six-dimensional metric is of the form
\be \label{e22}
ds_6^2=ds_4^2+\f{4\ell_p^2}{\Im \t} | d\hat\chi_m +\t d\hat\chi_e |^2,
\ee
with
\bea
&& d\hat\chi_e=d\chi_e-\f{e Q_e}{r}dt,  \nn\\
&& d\hat\chi_m=d\chi_m-\f{\f{Q_m}{e}-e Q_e\f{\th}{2\pi}}{r}dt,  \nn\\
&& \chi_{e,m} \sim \chi_{e,m}+2\pi.
\eea
The above uplifted metric is the solution of the six-dimensional Einstein gravity
\be
S=\f{1}{16\pi G_6} \int d^6x\sr{-G} R_6,
\ee
with $G_6=16\pi^2\ell_p^4$. The two extra dimensions form  a torus with the modular parameter $\t$. The torus is invariant under the modular group $SL(2,Z)$,
\bea \label{e57}
&& \t'=\f{d \t -c}{ -b \t+a},  \nn\\
&&
\lt( \ba{c} \chi_{e'} \\ \chi_{m'} \ea \rt)=
\lt(\ba{cc} a & b \\ c & d \ea \rt)
\lt( \ba{c} \chi_{e} \\ \chi_{m} \ea \rt),
\eea
with $a,b,c,d \in Z$ and $a d-b c=1$. Note that the uplift (\ref{e22}) cannot be done arbitrarily, and the uplifted metric has to be in accord with the quantization condition (\ref{quan}). From the geometry one could get the conserved angular momentum
\be
J_e=\f{Q_e}{e}-e Q_m \f{\th}{2\pi}=N_e, ~~~ J_m=e Q_m=N_m,
\ee
which must be integers. Thus the uplifted metric (\ref{e22}) is the only possible uplifting up to a possible $SL(2,Z)$ redefinition of the modular parameter $\t$ and the angles $\chi_{e,m}$ (\ref{e57}).

As we did in Section~2, we consider the extremal black hole and compute
\be
f_1=\f{1}{r_+}, ~~~ f_2=r_+, ~~~ f_3^{e}=\f{e Q_e}{r_+^2}, ~~~ f_3^{m}=\f{\f{Q_m}{e}-e Q_e\f{\th}{2\pi}}{r_+^2},
\ee
and thus
\be
f^{e}=\f{f_2 f_3^{e}}{f_1}=e Q_e, ~~~  f^{m}=\f{f_2 f_3^{m}}{f_1}=\f{Q_m}{e}-e Q_e\f{\th}{2\pi}.
\ee
 As the horizon area of the four-dimensional extremal black hole is $A_+=4\pi \ell_p^2 Q^2$,  we have the extremal version of the electric and magnetic CFT dual pictures
\bea
&& c_L^{e}=\f{3f^{e}A_+}{2\pi G_4}=6eQ_eQ^2, ~~~ T_L^{e}=\f{1}{2\pi f^{e}}=\f{1}{2\pi e Q_e},  \\
&& c_L^{m}=\f{3f^{m}A_+}{2\pi G_4}=6Q^2\lt( \f{Q_m}{e}-e Q_e\f{\th}{2\pi} \rt), ~~~
   T_L^{m}=\f{1}{2\pi f^{m}}=\f{1}{2\pi \lt( \f{Q_m}{e}-e Q_e\f{\th}{2\pi} \rt)}, \nn
\eea
They are in accord with the ones obtained in the thermodynamics method. After the redefinition of the modular parameter and the angles (\ref{e57}), one could get the $SL(2,Z)$ generated pictures in accord with the results before. Now the $SL(2,Z)$ duality could be understood as the geometric modular symmetry of the extra torus.

\section{Conclusion and discussion}\label{s6}

In this paper we further refined the thermodynamics method of setting up nonextremal black hole/CFT correspondence. The essential part of our improvement is to impose the quantization condition on the first laws. Physically, the quantization condition comes from the fact that the perturbation always carries integer units of angular momentum and/or charges. As a result,  the first laws of the black hole, encoding the response of the black hole with respect to the perturbation, should obey the quantization rule as well.  From the first laws of the outer and inner horizons we can in general have
\be \label{e61}
d N=T_L^N d S_L -T_R^N d S_R,
\ee
with $N$ being an integer quantized charge and all other charges being kept invariant. Then the temperatures of the $N$ picture CFT dual is $T_{R,L}^N$, and the central charges could be derived using the Cardy formula $c_{R,L}^N=\f{3}{\pi^2}\f{S_{R,L}}{T_{R,L}^N}$. On the other hand, taking (\ref{e61}) as the first laws for the underlying CFT requires reasonably the quantization condition. Certainly how to understand (\ref{e61}) in the underlying CFT is an important issue.

We investigated the holographic descriptions of various RN black holes via the thermodynamics method. As showed in \cite{Chen:2012mh}, the relation $T_+S_+=T_-S_-$ could be taken as the criterion to see if a black hole may have a CFT dual in the Einstein gravity.  We found that $T_+S_+=T_-S_-$ holds for RN black holes in all dimensions, which means that RN/CFT correspondence could be generalized to all dimensions, and  $T_+S_+=T_-S_-$ breaks down for RN-AdS black holes, which forbids us finding their CFT duals. Moreover, we tried to set up CFT duals explicitly for RN black holes in various dimensions by using the first laws at the outer and inner horizons. It turned out that all the pictures we found are in agreement with the ones read from conventional ASG analysis and the hidden conformal symmetry.

It is remarkable that the refined thermodynamics method resolve the puzzle on the ambiguity in determing the central charges of CFTs dual to the RN black holes. Starting from the first laws (\ref{e61}), there is no ambiguity in deciding the central charges. For example, for the four-dimensional RN black hole, the CFT dual has central charges $c=6eQ^3$, where $e$ is the unit electron charge.
The quantization condition is actually reflected in the facts that  the angular momentum along the extra circle in the uplifted spacetime must be quantized, and in the discussion of hidden conformal symmetry the identified angular quantum number should be integer-valued as well.  These key points have been ignored in the literature.

Besides the RN black holes in various dimensions, we also discussed 4D dyonic RN black holes and found the novel magnetic picture. This picture have not been discussed in the literature, partially due to the difficulty in deciding its central charge. The different kinds of pictures could be most easily read from the refined thermodynamics, while can be also seen from other points of view. For example, to read the magnetic picture from the hidden conformal symmetry, we had to consider the probe scalar field with magnetic charge scattering off the black hole. In the minimal coupling, while the electric charge couples to the gauge potential, the magnetic charge couples to the dual gauge potential by electromagnetic duality. Such a coupling indeed gives us a consistent magnetic picture. For convenience we call the electronic and magnetic pictures (1,0) and (0,1) picture respectively. As shown in (\ref{e28}) there are other pictures generated by $SL(2,Z)$ group. In general we could obtain a $(a,b)$ picture for every coprime integers $a,b$. This is a kind of duality among different CFT theories. How to understand this duality is an interesting issue. We tried to understand the dyonic black hole by relating it to a $U(1)^2$ two-charged black hole, from which we may read the central charges from ASG analysis of an uplifted six-dimensional metric and more remarkably we may interpret the underlying $SL(2,Z)$ symmetry as the modular group of the extra torus. However, it would be much better to investigate the symmetry from the dyonic black hole itself. In \cite{Chen:2012pt}, it has been shown that the $SL(2,Z)$ duality originates from the electromagnetic duality of the theory. The basic point is that the dyonic black hole spacetime is invariant under the electromagnetic duality, even though the charges of the black hole and the gauge potential have to be transformed. Therefore it is feasible to describe the same black hole in different $SL(2,Z)$-related theory. Note that the notion of electromagnetic duality is new in the context of 4D Einstein-Maxwell theory, but is well-known in string theory \cite{Cvetic:1995yp,Cvetic:1995yq}.

The effectiveness and rebustness of  the thermodynamics method is remarkable. It not only allows us to read the temperatures and the central charges of dual pictures, it also helps to determine the the frequencies, the charges and the chemical potentials of dual operators if one consider a perturbation, in exact agreement with the ones got from the low frequency scattering. The underlying pictures of the thermodynamics method and the hidden conformal symmetry are different. Consider the situation that a particle falling into the black hole. The particle carries energy, possibly also an angular momentum and various charges. The response of the black hole with respect to the infalling particle is encoded in the thermodynamics laws, both at the outer and inner horizons. There is a dual operator in CFT corresponding to such a perturbation. The thermodynamics method shows how to read various information on dual CFT. The different pictures can be obtained by considering the response of the black hole with respect to different kinds of perturbations. For example, in the 4D dyonic Kerr-Newman case, if the perturbation carries only an angular momentum, the response gives us the $J$ picture, while if the perturbations carries only an electric (magnetic) charge, then the response gives us the electric (magnetic) picture. In general, if the perturbation carries all the quantum charges, it may lead to a new picture, whose information could be obtained by $SL(3,Z)$ transformation acting on the elementary pictures. On the other hand, a probe scattering off the black hole, especially at the low frequency limit in the near region, can tell us the information of the black hole as well. In a quite similar way, different probe could read out different dual pictures as we showed.

In all the cases we studied the thermodynamics method is more powerful than the hidden conformal symmetry in the sense that even for the cases in which the hidden conformal symmetry is not easy to find, say the black ring case  \cite{Chen:2012yd}, the thermodynamics method can still give us consistent pictures. Nevertheless, it would be interesting to understand better the relation of these two different methods \cite{Chen:2013rb}.

The accumulated evidence supports that the thermodynamics method is universal to decide the dual holographic picture of black hole. The thermodynamics of outer and inner horizon seems encode all the information of the dual CFT. It would be important to have a better understanding of the  physics underlying this method.

\vspace*{10mm}
\noindent {\large{\bf Acknowledgments}}\\
The work was in part supported by NSFC Grant No.~10975005,~11275010. JJZ was also in part supported by Scholarship Award for Excellent Doctoral Student granted by Ministry of Education of China.
\vspace*{5mm}




\end{document}